Article

# Microscopic Characterization of Individual Submicron Bubbles during the Layer-by-Layer Deposition: Towards Creating Smart Agents

Riku Kato [1,2], Hiroshi Frusawa [2,*]

[1] Faculty of Pharmaceutical Science, Nagasaki International University, Huis Ten Bosch, Nagasaki 859-3298, Japan; E-Mail: riku@niu.ac.jp

[2] Research Center for Materials Science and Engineering, Kochi University of Technology, Tosa-Yamada, Kochi 782-8502, Japan

* Author to whom correspondence should be addressed; E-Mail: frusawa.hiroshi@kochi-tech.ac.jp.

**Abstract:** We investigated the individual properties of various polyion-coated bubbles with a mean diameter ranging from 300 to 500 nm. Dark field microscopy allows one to track the individual particles of the submicron bubbles (SBs) encapsulated by the layer-by-layer (LbL) deposition of cationic and anionic polyelectrolytes (PEs). Our focus is on the two-step charge reversals of PE-SB complexes: the first is a reversal from negatively charged bare SBs with no PEs added to positive SBs encapsulated by polycations (monolayer deposition), and the second is overcharging into negatively charged PE-SB complexes due to the subsequent addition of polyanions (double-layer deposition). The details of these phenomena have been clarified through the analysis of a number of trajectories of various PE-SB complexes that experience either Brownian motion or electrophoresis. The contrasted results obtained from the analysis were as follows: an amount in excess of the stoichiometric ratio of the cationic polymers was required for the first charge-reversal, whereas the stoichiometric addition of the polyanions lead to the electrical neutralization of the PE-SB complex particles. The recovery of the stoichiometry in the double-layer deposition paves the way for fabricating multi-layered SBs encapsulated solely with anionic and cationic PEs, which provides a simple protocol to create smart agents for either drug delivery or ultrasound contrast imaging.

## 1. Introduction

Microbubbles designate colloidal gas particles in aqueous media possessing diameters of 1 to 100 micrometers [1–6]. The dispersed microbubbles have been found to be unstable due to the effects of their surface tension: the microbubbles tend to leak into aqueous solutions due to a large Laplace pressure that drives the dissolution of the bubbles [7,8], and they need to be covered with a variety of stabilizing shells [1–23]. The diffusion of the gas can be slowed by coating the gas core, thereby increasing the lifetime to a few months [4,5,15]. Recently, we have also found that smaller dispersions of gas particles, which will be referred to as submicron bubbles (SBs) below, can exist for more than hours even in the absence of stabilizers [1,24–32].

The colloidal bubbles, including SBs as well as microbubbles, have many unique properties, such as a low rising velocity, a friction reducing effect, and a high pressure inside of the bubble. Thanks to these novel features, much attention has been paid to micron and/or submicron bubble suspensions in various fields [1]. For example, colloidal bubbles were used for water treatment in the engineering and industrial fields [1,21,31–34]. Effective treatment methods for wastewater with microbes, including cyanobacteria, were investigated, thereby achieving the removal of negatively charged bacteria from water using positively charged microbubble complexes [21]. The low gas permeability of encapsulated bubbles also facilitates an efficient trapping of functional gas inside solutions, which is useful for agricultural and green-technological applications, such as the industrial food applications [33] and the removal of residual pesticides from leaves using ozone microbubbles [34].

In terms of biomedical use, it should be noted that microbubbles are similar to red blood cells in size and are expected to display a similar rheology. A wide variety of the biomedical applications include the exploitation of the colloidal bubbles as agents not only for contrast ultrasound imaging but also for targeted drug and gene delivery [1–3,6,11,12,14–16,20,35–43]. The efficiency of delivering a drug to the specific target region has been improved by focusing an ultrasound field to destroy the microbubble carriers in order to release drugs or therapeutic genes. Furthermore, coating the gas core with a shell is significant for loading the drug or DNA molecules as well as for stabilizing the colloidal bubbles. These shells have been fabricated using various materials, such as proteins and lipids, and with neutral and charged surfactants, yet the thin shell on the gas-filled core itself cannot be used as an efficient therapeutic vehicle. The loading capacity of the bubble for drug or gene therapies must be further increased [3,6,11,12,35,36].

One of the most promising shells for this purpose is the polyelectrolyte (PE) multi-layer formed by the layer-by-layer (LbL) method [2,9–12,15,17,18]. The LbL technique is essentially based on charge reversal phenomena whereby colloidal particles are overcharged by adding oppositely charged PEs [26,27]. There are some potential advantages to LbL encapsulation. First, the overall loading capacity of microbubble surfaces can be increased due to the electrostatic sandwiching of anionic DNA between cationic layers [11,12]. Additionally, the polymeric shells facilitate the ability to impart a good degree of biocompatibility or biodegradability as well as additional functionality, such as molecular recognition.

Nevertheless, most of the first coverage on the uncoated microbubbles has been performed using non-polymeric materials including lipids and proteins. While there was a previous report that used $CO_2$ microbubbles for the preparation of multi-layers of pure PEs [15,17,18], it has been suggested that uncoated microbubbles of air or oxygen have some difficulty in implementing the charge reversal caused by cationic PEs (polycations) that violates the electrical stoichiometry [26,27], though a wide variety of colloid-PE complexations have almost satisfied the stoichiometry at the isoelectric point [44]. We have ascribed the anomalous phenomena to the anions on the surface of unmodified air or oxygen microbubbles that arise from the adsorption of hydroxyl ions on the gas surface [26,27], except for in strong acid solutions [1,45,46]; however, the details remain to be addressed, and we need to investigate the monolayer deposition regarding various species of polycations and compare the first and second charge reversals due to the depositions of mono-and double-layers, respectively.

In this paper, we aim to experimentally clarify whether the mono-and double-layer deposition processes take different courses or not, partly because such comparison serves to judge the validity of previous model [27] that theoretically explains the non-stoichiometric anomaly found in the mono-layer depositions [26,27]. To this end, we performed *in situ* observations of SBs using dark field microscopy during the LbL deposition procedures followed by gradually adding polycations (mono-layer deposition) or polyanions (double-layer deposition). Our focus is on tracking the trajectories of individual gas particles that experience either Brownian motion or electrophoresis without forming aggregates. It is therefore required that the uncoated colloidal bubbles should not disappear with an experimental time, which the bare SBs meet as mentioned above.

Thus, we analyzed the trajectories of SBs in extremely dilute suspensions so that coagulation might be suppressed considerably. With the electrophoresis measurements, our particular concern is the isoelectric point where the electrophoretic mobility µ is negligible on average due to the electrical neutralizations of either the mono-layered SBs (M-SBs) covered with polycations or the double-layered SBs (D-SBs) covered with polyanions on the surface of mono-layered SBs. The isoelectric point is specified by the critical monomer density $C_{cm}$ of the cationic or anionic PEs added to cancel the effective charges of the PE-SB complex particles. We will determine $C_{cm}$ in various complex solutions by forming M-SBs or D-SBs to investigate whether the electrical stoichiometry is satisfied at the isoelectric points. Furthermore, we have corroborated that the Brownian particles tracked for the electrophoresis measurements are neither aggregates nor dust.

This paper is organized as follows. The next section is the experimental section, where we will describe the details regarding the analysis of particle trajectories as well as the materials and experimental systems that were utilized in this research. The next section includes the results and discussion and will consist of four parts. In the first part, we will discuss the size distribution of bare SBs with no shells due to the absence of stabilizers, such as PEs, in the suspension. In the second part, we will investigate the pH-dependence of µ and the associated zeta potential for bare SBs, and will compare the variations in µ with polycations, polyanions and neutral polymers added, respectively. In the third section, we will demonstrate the different behaviors between the mono- and double-layer deposition processes by focusing on the stoichiometry in terms of the total charge ratio at the isoelectric points of M-SBs and D-SBs. The fourth part will provide the size distributions of M-SBs and D-SBs for a comparison. Finally, the last section will contain concluding remarks on the implications of creating smart agents for either drug delivery or ultrasound contrast imaging.

## 2. Experimental

*2.1. Materials*

The SB particles made of air were produced by a fine bubble generator (OM4-GP-040, Aura Tec, Kurume, Japan) in water that had been deionized and filtered using a 200-nm-pore-size filter (Minisart, Sartorius, Hannover, Germany) in advance. The SB suspensions were diluted to fix the SB density at approximately 3 fM by adding salt-free water or PE solutions. A previous study of the same SBs evaluated by the comparison of the micrographs of SBs and silica colloids with known concentrations found that the density of 3 fM was maintained for several hours, at least [26], which was sufficient to finish the measurements while suppressing the coagulation between SB particles surrounded by various polymers.

Cationic PEs added to the bare SB suspension were poly (diallyldimethyl ammonium) chloride (pDADMAC), poly-L-lysine (PLL), and poly(allylamine hydrochloride) (PAH). The distributions of the weight-averaged molecular weight ($M_w$) were in the order: $M_w$ < 100,000 (pDADMAC, Sigma, St. Louis, MO, USA), $M_w$ = 150,000–300,000 (long PLL (L-PLL), Sigma), $M_w$ = 70,000–150,000 (short PLL (S-PLL), Wako), $M_w$ = 56,000 (long PAH (L-PAH), Sigma), and $M_w$ = 17,000 (short PAH (S-PAH), Sigma). We also used polyethylene glycol (PEG, $M_w$ 5000–7000, Fluka, Tokyo, Japan) as a neutral polymer and anionic PEs of sodium poly(styrenesulfonate) (NaPSS, Pressure Chemical, Pittsburgh, PA, USA) with narrowly distributed molecular weights (polydispersity index < 1.5): $M_w$ = 990,00 (L-NaPSS), $M_w$ = 208,000 (M-NaPSS), and $M_w$ = 8,000 (S-NaPSS). We passed all species of the polymer solutions through a 200-nm-pore-size filter (Minisart, Sartorius) prior to mixing with the salt-free SB suspension to eliminate aggregates or dusts in the polymer solutions with no SB particles added. To adjust the solution pHs, we used a phosphate buffer (Wako) as well as HCl and NaOH (Sigma).

Mono-layer deposition processes were investigated by performing Brownian and electrophoretic measurements of various SB-polycation suspensions, each of which was prepared by mixing a bare SB suspension and a polycation solution with its monomer density varying in a range of $C_m$ < 5 μM. Polyanions for double-layer deposition, on the other hand, were added to a positively charged M-SB suspension which had been prepared beforehand in an S-PLL solution with its monomer density fixed at 2 μM. Similarly to the investigation of mono-layer formation, different courses of double-layer deposition were compared using mixtures of the M-SB dispersion and polyanion solutions with its monomer density changed within a range of $C_m$ < 5 μM.

*2.2. Experimental Setup*

The concentration of SB particles was fixed at a low density of 3 fM for all experimental conditions. We characterized the varying solution conditions using a pH and conductivity meter (Seven Easy, Mettler Toledo, Columbus, OH, USA). The dispersed SBs undergoing Brownian motion or electrophoresis were detected at 25 °C via dark field microscopy using a zeta potential analyzer (Zeecom, Microtech Nition, Funabashi, Japan). Video observations of the migrating particles (see Movie S1 in Supplementary Materials) enable one to measure the individual velocities by tracking the particles, respectively; this method has been referred to as the microscopic electrophoresis method. We used a rectangular cell with dimensions of 1-cm in height ($2h$ = 10 mm), 0.75-mm in depth ($2d$ = 0.75 mm), and 9-cm in length, which is equal to the electrode gap length.

As control experiments, we prepared a variety of solutions without the use of a bubble generator for observing, in the absence of SBs, a salt-free solvent of deionized water, a polycation solution with a monomer density of $C_m$ = 2.0 μM, and a mixture of polyanions and polycations whose monomer densities were set to be 2 μM and 3 μM, respectively. These solutions were filtered and the monomer density adopted was much lower than the overlapped density beyond which PE solutions can be regarded as semidilute solutions [47]. Accordingly, dark field microscopy of these bubble-free solutions was unable to detect any particles undergoing Brownian or electrophoretic motions. The consistently null results in the control observations conversely reveals that the observed particles in the presence of bubbles are SBs and/or PE-SB complex particles.

The number of SB particles tracked and analyzed in a set of Brownian or electrophoretic experiments was fixed at fifty, and we repeated three to five 50-particle measurements at each experimental condition. Thanks to careful filtering prior to mixing SB suspension and a PE solution, there were only a few aggregates that were distinguishable due to the obviously high intensity of scattered light while collecting trajectory data of 50 particles. Therefore, the microscopic observations can exclude the aggregates from the analyses, other than the dynamic light scattering (DLS) method that necessarily analyzes an integrated intensity of all scattered light.

*2.3. Analysis*

The video analysis of Brownian particles provides the mean square displacement (MSD), which is proportional to a time interval $t$:

$$\langle \mathbf{r}^2(t) \rangle = 2dDt \tag{1}$$

where $\mathbf{r}(t)$ denotes the particle displacement and $D$ the diffusion constant of the SBs. We adopted $d = 2$ as the spatial dimension because $\mathbf{r}(t)$ lies in a two-dimensional plane in the measurement of the MSD in the movies. The hydrodynamic diameter $2a$ of the SB is thus evaluated from combining the diffusion constant $D$ obtained from Equation (1) and the Stokes-Einstein relation:

$$D = \frac{k_\mathrm{B} T}{6\pi\eta a} \tag{2}$$

with the viscosity of water and the thermal energy denoted by $\eta$ and $k_\mathrm{B}T$, respectively.

We evaluated the inherent electrophoretic drift velocity $v_\mathrm{d}$ from fitting the theoretical velocity profile $v(x)$ to the mean velocity distribution obtained experimentally, with the average of fifty particles tracked using dark field microscopy. In the case of the rectangular housing, $v(x)$ is given by [48]

$$v(x) = v_\mathrm{d} + v_\mathrm{eo} \left[ 1 - 3 \left\{ \frac{1-(x/d)^2}{2 - 384\, d/(\pi^5 h)} \right\} \right], \tag{3}$$

where $v_\mathrm{eo}$ denotes the electroosmotic velocity in the vicinity of the cell walls.

For all of the experiments, we have verified a range of an effectively applied electric field $E$ that satisfies the linear relationship of electrophoretic velocity $v_\mathrm{d}$ *versus* $E$; therefore, the slope of the fitting line provides the electrophoretic mobility $\mu$.

There are various equations relating the zeta potential $\zeta$ to $\mu$. From them, we used Henry's equation, which has been found to cover a wide range of solution conditions [49]:

$$\mu = \frac{\varepsilon_\mathrm{r} \varepsilon_0 \zeta}{\eta} f(\kappa a) \tag{4}$$

where $\varepsilon_\mathrm{r}$ and $\varepsilon_0$ are the permittivities of the water and the vacuum, respectively, and $f(\kappa a)$ denotes Henry's coefficient as a function of the length ratio, $\kappa a$, of the hydrodynamic radius $a$ to the Debye length $\kappa^{-1}$ which was determined from the measured conductivity. Thanks to the analytical form derived by Oshima *et al.,* [49], the zeta potential ξ directly provides the mean surface charge density σ, or the average effective charge $Z_B = 4\pi a^2 \sigma$ that an SB carries.

We use the charge ratio, $\alpha = Q_p/Q_B$, between the added charges $Q_p$ of the ionic groups on the PE chains in total and the sum of the charges of the SBs, $Q_B$, as indicators of the aforementioned stoichiometry. Fixing the volume at 1 L for brevity, $Q_B$ is given by $Q_B = eZ_B C_B$, using the density, $C_B$, of the SBs, whereas we have $Q_p = eC_{cm}$ with $e$ denoting the elementary charge and the supposition that every constituent monomer used herein has a monovalent ionic group. Accordingly, α can be determined by $\alpha = C_{cm}/(Z_B C_B)$, where $C_B$ is set to be a constant under the conditions in generating the SBs.

## 3. Results and Discussion

*3.1. Evaluating Hydrodynamic Diamters of Bare SBs*

Figure 1a shows a typical result of the MSD *versus* time lag *t* for a bare SB in salt-free water at 25 °C. In Figure 1a, the line delineates the fitting results of Equation (1). It was determined from Equation (1) that the slope of the fitted line in Figure 1a was proportional to the diffusion constants of a bare SB in a two-dimensional plane. We further evaluated the hydrodynamic diameter, 2*a*, from Equation (2), providing that 2*a* = 250 nm for the targeted SB in Figure 1a.

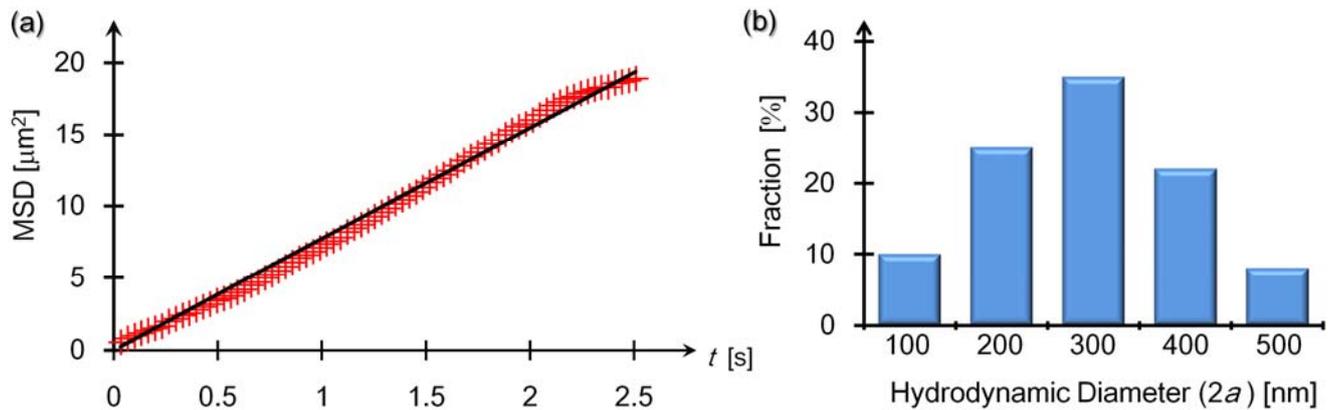

**Figure 1.** Time-interval (*t*) dependence of the MSD of a bare SB (**a**) and histogram representation of the diameter distribution of 50-particles that were recognized as bare SBs (**b**). (**a**) The line delineates Equation (1) that was fitted to the measured MSD (red crosses). Error bars lie within symbols. (**b**) A percentage denotes a mean fraction of 50 particles which depends on a range of hydrodynamic diameter. For example, the label "100" in the transverse axis represents the range $100 \leq 2a < 200$.

Repeating the MSD measurements as shown in Figure 1a, we created the histogram in Figure 1b that represents the distribution of the hydrodynamic diameter for the suspensions of bare SBs. The mean diameter 2*a* of the bare SBs was 2*a* = 320 nm, which agrees with the previous results obtained from the DLS method [25,28,31]. Incidentally, we can estimate the upper limit of the bubble density $C_B$ using the mean radius, *a* = 160 nm, in addition to the injected air volume $V_a$ in 1 liter of the suspensions: we have $C_B \leq V_a/(N_A v_B) \approx 6$ fM consistently with the actual density adjusted to $C_B = 3$ fM (see Section 2.1), where $N_A$ and $v_B = (4\pi a^3)/3$ denote Avogadro's number and the volume of a single SB, respectively.

*3.2. pH Effects on Bare SBs and M-SBs*

Following the treatment based upon Henry's Equation (4), we calculated the zeta potentials of the bare SBs from the obtained values of µ for the various solvents with different pHs (Table 1). It has been explained that bare microbubbles carry charges due to the adsorption of either a hydroxyl ion $OH^-$ or a hydronium ion $H_3O^+$, which naturally caused the pH-induced charge reversal. The pH dependence in Table 1 indicates that lowering the pH converts anionic bubbles into cationic ones, which is similar to that of the microbubbles that were previously reported [45,46]. We also find from Table 1 that the absolute value of the zeta potential increases when increasing the solution pH. The results at both the low and high pHs were consistently explained by the existence of SBs at pH 7.1 that were negatively charged due to the adsorption of the $OH^-$ that was dissociated from the water molecules.

**Table 1.** Mean zeta potentials of bare SBs at various pHs.

| Salt conditions | pH | Zeta potential (mV) |
|---|---|---|
| HCl (0.1 mM) | 4.1 | +14 |
| Buffer (1 mM) | 7.1 | −38 |
| Salt free | 7.1 | −40 |
| NaOH (0.1 mM) | 10.4 | −50 |

Figure 2 shows variations of the elecrophoretic SB-mobilities, µ, due to the addition of a polycation (S-PLL), a polyanion (L-NaPSS), and a neutral polymer (PEG). In Figure 2, we observe two $C_m$-dependencies of µ, which is indicative of the charge reversals that were induced by the addition of the oppositely charged polycations to the negatively charged SB suspensions in the buffer solution and with the addition of 0.1 mM-NaOH. The resulting pHs fell within the range of 7.0 to 7.2 in the buffer solution and were maintained at a pH between 10.0 and 10.5 in the 0.1 mM-NaOH solutions. The mobility of the SB particles at a pH of 4.2 in 0.1 mM-HCl solutions, on the other hand, remained positive irrespective of $C_m$ though this is not included in Figure 2.

The contrasted behaviors of µ at different pHs strongly suggest that the charge reversals in Figure 2 were not induced by the pH changes, but were driven by the electrostatic attractions between the anionic SBs and the polycations. To verify the electrostatic mechanism, we also performed control experiments that investigated the variations in µ in the presence of neutral polymers or similarly charged polyanions. Figure 2 further shows that SBs with the addition of neutral or anionic polymers possessed negative mobility as well as that of the bare SBs, thereby revealing that the electrostatic attractions are indispensable for inducing the present charge reversals due to the addition of the oppositely charged polycations.

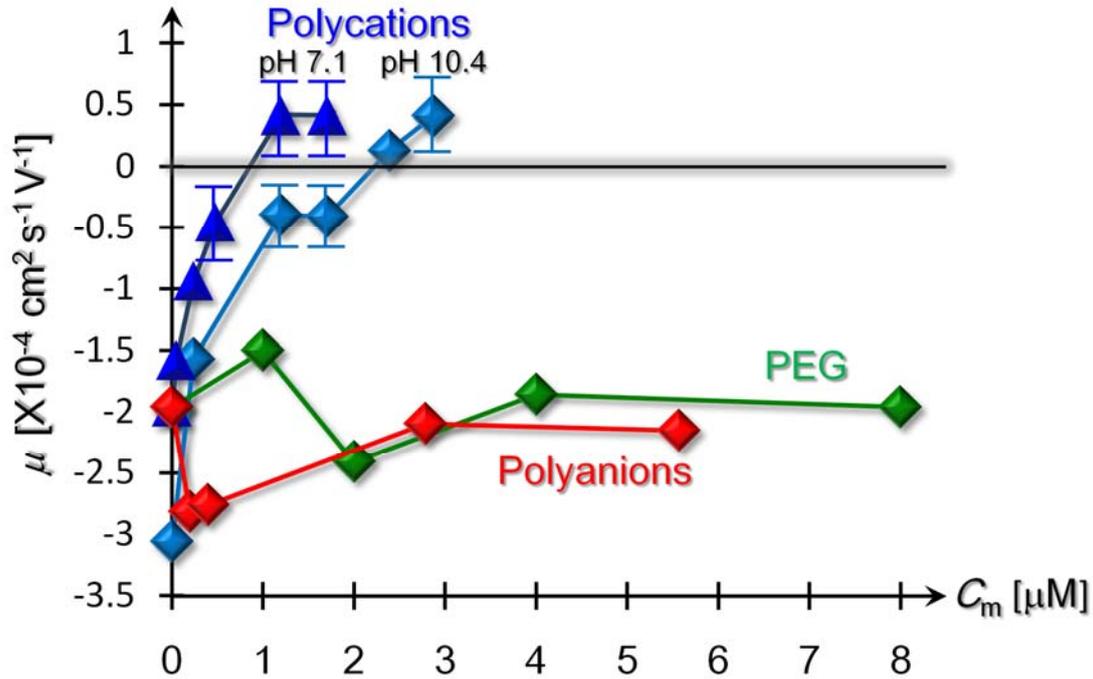

**Figure 2.** Changes of electrophoretic mobilities due to the addition of polycations (S-PLL) at pHs of 7.1 and 10.4, neutral polymers (PEG) and polyanions (L-NaPSS). Error bars lie within symbols if unspecified.

From Figure 2, we can evaluate the isoelectric point, specified by a critical monomer density $C_{cm}$, at which μ vanishes due to the electrical neutralizations of the SBs, and beyond which the sign of μ is reversed. Figure 2 provides the different ranges of 0.5 μM < $C_{cm}$ < 1.2 μM and 1.7 μM < $C_{cm}$ < 2.4 μM where neutral and basic solvents were used, respectively. Meanwhile, it should be noted that, as observed in Table 1, the zeta potential increases with an increase in pH, arising from the difference of μ at $C_m$ = 0 μM in Figure 2, which indicates that the SBs carry more negative charges in the basic solvents than in neutral solutions. Combining these results, we can ascribe this delay of the charge reversals with the increase of pH to the increasing number of anions adsorbed onto the SB surfaces.

*3.3. Comparison of the Charge-Reversal Phenomena due to the Mono-and Double-Layer Depositions.*

Figure 3a depicts the $C_m$ dependencies of μ in five kinds of salt-free polycation solutions. It should be noted that no salt was added in Figure 3a, as opposed to Figure 2; nevertheless, the resulting pHs of the salt-free solutions were maintained in the range of 6.6 to 7.4 overall for the monomer density adopted. Despite the small variance in the pHs, we observed from Figure 3a that $C_{cm}$ ranges from the minimum of $C_{cm}$ < 1 μM to the maximum of $C_{cm}$ > 2 μM, and the diversity of $C_{cm}$ is as large as the aforementioned difference of $C_{cm}$ at pH values of 7.1 and 10.4. Such variety in the $C_{cm}$ arises from the differences of polycation species or the molecular weight $M_w$ (equivalent to the chain length) of the same species. First, the species dependency of $C_{cm}$ suggests that the effective dissociation of the ionic groups was affected by the constituent molecules due to various factors such as the counterion condensation, even though the degree of dissociation was chemically indiscernible. Actually, the difference of $C_{cm}$ associated with the distinction of the constituent molecules has also been reported in previous charge reversal phenomena of conventional colloids, such as silica particles and polystyrene colloids [45].

More attention should be paid to the other variations of $C_{cm}$ for the identical species of PLL and PAH. Figure 3a shows that $C_{cm}$ increased when the polycation was longer, indicating that the isoelectricity of the individual SBs is not governed by the electrical stoichiometry of the anionic SBs and the polycations other than that for the previous charge reversal phenomena of usual colloids [26,27,45].

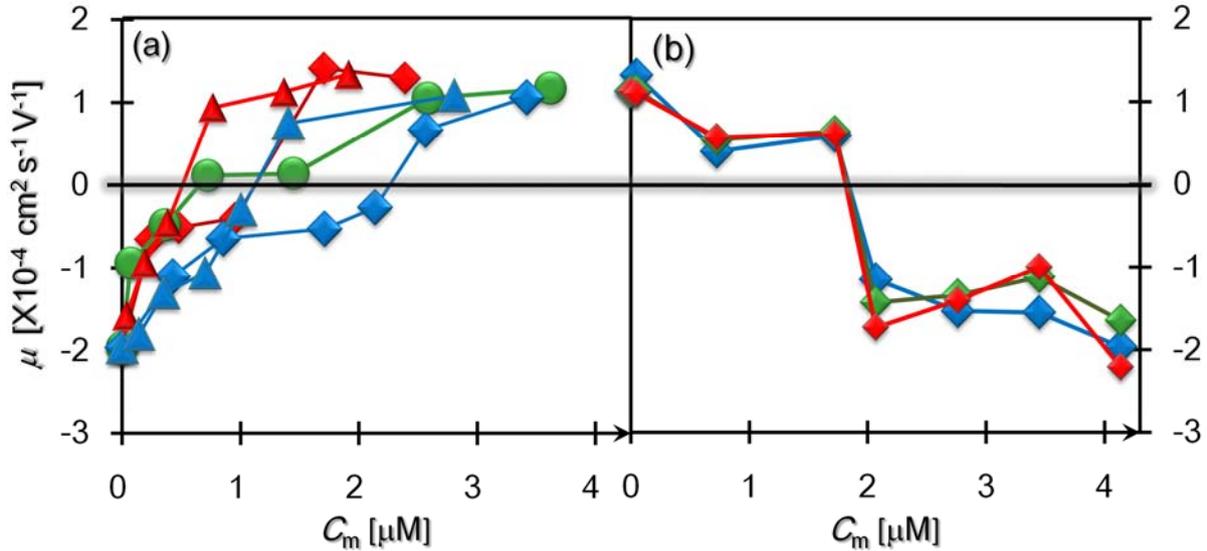

**Figure 3.** Charge reversals due to the formations of a polycationic mono-layer (**a**) and of a polyanionic and polycationic double-layer (**b**). (**a**) Electrophoretic mobilities of the M-SBs where various species of polycations (S-PLL (red triangles), L-PLL (red diamonds), pDADMAC (green circles), S-PAH (blue triangles), and L-PAH (blue diamonds)) were added. (**b**) Double-layered SB complexes formed by the addition of polyanions (S-NaPSS (red), M-NaPSS (green) and L-NaPSS (blue)) with three kinds of molecular weights to the positively charged M-SBs. Error bars lie within symbols.

We introduce the charge ratio $\alpha = C_{cm}/(Z_B C_B)$, defined in Section 2.3, in order to determine the extent to which the conventional stoichiometry was violated. The effective charge number $Z_B$ per SB can be obtained from $\mu$ at $C_m = 0$ as described above, providing that $Z_B \approx 400$ as reported previously [26,27]. Incidentally, as described in Section 2.1, we set that $C_B = 3$ fM. Table 2 lists the various values of $\alpha$, which have been obtained from $C_{cm}$ in Figure 3a. The above dependencies of $C_{cm}$ on the polycation species and $M_w$ were reflected by the variations of these results in Table 2. It is far more anomalous that all of the results of $\alpha$ are on the order of $10^4$, which is much larger than the stoichiometric condition of $\alpha = 1$.

**Table 2.** The critical monomer-concentrations of the various polycations added for overcharging, and the charge ratio at the charge reversal.

| Polycations | $C_{cm}$ [μM] | Charge ratio $\alpha$ ($\times 10^4$) |
|---|---|---|
| pDADMAC | 0.4–0.7 | 6–12 |
| L-PLL | 1.0–1.7 | 16–28 |
| S-PLL | 0.4–0.8 | 6–13 |
| L-PAH | 2.1–2.6 | 35–42 |
| S-PAH | 1.0–1.4 | 16–23 |

Next, Figure 3b displays three kinds of charge reversals induced by the addition of polyanions (NaPSS) to the positively charged M-SBs, where we used the same type of M-SBs equally formed in 2 μM-solutions of S-PLL. The added polyanions with short, medium and long chain lengths (S-, M- and L-NaPSS) possess narrow distributions of their molecular weights, as described in Section 2.1. In these experiments, an overwhelming excess of free polycations surrounding M-SBs had not been removed, and most of the 2-μM polycations were uniformly dispersed instead of covering the M-SBs, as inferred from the tremendous values of α in Table 2.

In contrast to the above diversity of $C_{cm}$ (see Figure 3a), the $C_m$-dependencies of μ in Figure 3b were quite similar; these results exhibited not only the coincidence with the $C_{cm}$, but also the identical reduction in μ prior to the isoelectric point. From Figure 3b, we can find the location of $C_{cm}$ as before, which is in a relatively narrow range of 1.8 μM < $C_{cm}$ < 2.0 μM. Since the number density of M-SBs is equal to that of bare SBs as confirmed from microscopic observations, the present range of $C_{cm}$ remains anomalous in terms of α. Considering the existence of free polycations, however, the anion density, which is equally close to 2 μM irrespective of molecular weights, is balanced with the concentration of free and deposited cations in total, indicating that the electrical neutrality of the entire system is satisfied at the isoelectric point of the double-layer deposition.

*3.4. Particle Size Distributions with Polyelectrolytes Added.*

Our preceding discussions in Section 3.3 have been made supposing that the particles tracked for evaluation of the individual mobilities were PE-SB complex particles. Indeed, the validity of the above analyses during the deposition processes is corroborated by continuous changes such that the mobilities gradually vary with an increase in $C_m$ as shown in Figures 2 and 3, and yet it remains to be verified whether the measured particles are actually SB-included ones. Hence, we further analyzed the dark field micrographs regarding three kinds of Brownian particles as follows: negatively charged M-SBs (NM-SBs) with polycations (S-PLL) of $C_m$ = 0.2 μM added (corresponding to the second red triangle from the left in Figure 3a), positively charged M-SBs (PM-SBs) with polycations (S-PLL) of $C_m$ = 2 μM added (the sixth red triangle from the left in Figure 3a), and negatively charged D-SBs (ND-SBs) at $C_m$ = 2.8 μM of polyanons (S-NaPSS) which are represented by the fifth red diamond from the left in Figure 3b.

First, dark field microscopy at any of the above suspensions detected no apparent change in SB density (see also Movie S1 in Supplementary Materials), compared with the bare SB concentration. We also investigated variation in the size distribution of the Brownian particles using Equation (2) as before. Because the added concentrations of PEs are so low, as mentioned in Section 2.1, that viscosity measurement cannot provide a relevant increase in PE-solution viscosity [47], we used the same water viscosity at 25 °C as that of bare SB suspensions when the hydrodynamic diameter 2$a$ was determined from Equation (2); actually, applying such an evaluation of particle size to that of negatively charged SBs with neutral polymers or polyanions added (green and red diamonds in Figure 2), we obtained mean diameters in a reasonable range of 300 to 400 nm, which is close to that of bare SBs.

Figures 4a–c show the diameter distributions of the NM-SBs, PM-SBs, and ND-SBs, from which we can determine mean diameters of 310, 420, and 400 nm, respectively. Comparison between the histograms of Figure 1b, Figure 4a–c indicates not only a similarity in the distributions between Figures 1b and 4a, but also a resemblance between Figures 4b,c, so that they can be classified into two groups: one (Figure 2b and Figure 4a) and the other (Figures 4b,c). The difference is that both distributions of Figure 4b,c have longer tails extended to a larger size than those of bare SBs (Figure 1b) and NM-SBs (Figure 4a); correspondingly, the mean diameters of overcharged particles (PM-SBs and ND-SBs) are longer than that of bare SBs by approximately 100 nm. It should be also noted that, despite the distribution similarity, the surrounding PE solutions are in distinct states: while PM-SBs are surrounded by an excess amount of polyanions, ND-SBs exist in complex PE solutions of polycations and polyanions.

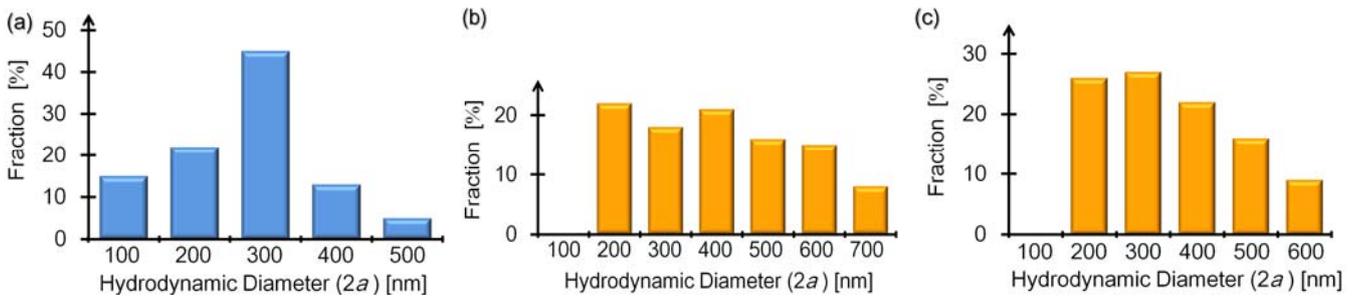

**Figure 4.** Histogram representations of diameter distributions regarding polycation-deposited SBs (NM-SBs (**a**) and PM-SBs (**b**)) and polyanion-polycation-deposited SBs (ND-SBs (**c**)). Both of the transverse and longitudinal axes denoted the same as those of Figure 1b where the distribution of bare SBs has been shown.

Combining these features of both the number density and the diameter distribution of observed particles in PE solutions, it is verified that the charge reversal phenomena shown in Figures 2 and 3 are due to the LbL depositions of polyelectrolytes on the surface of submicron particles. In other words, Figures 2 and 3 evidently demonstrate the overcharging phenomena of PE-SB complex particles, and the broader distributions of overcharged particles as seen from Figures 4b,c were caused not by bubble-free complexes of polyanions and polycatios, but by clustering, such as dimerization, of PE-SB complex particles whose shells were mono-layered or double-layered.

## 4. Conclusions

In conclusion, the use of dark field microscopy has clarified the properties of the individual SB particles instead of the averaged results of the SB suspensions. It was experimentally significant that the initial size of the SBs was maintained for more than several hours [24–31]. Thanks to the existence of the stable SBs, we were able to compare the detailed processes of the mono- and double-layer depositions by the addition of polycations and polyanions, respectively, though such a long lifetime of the uncoated SBs remains to be rationalized. Our measurements were devoted to extensively tracking the trajectories of the Brownian and electrophoretic motions (see again Movie S1 in Supplementary Materials). It should be noted here that our concern has not been with the size distribution of all particles existing in the suspensions, but rather with the sizes of the particles that undergo electrophoreses. The dark field microscopy method can meet this requirement, other than the DLS technique. We have found from the MSD analysis regarding Brownian particles that the major part of the clusters consisted of a few SBs due to the suppressed emergence of the huge aggregates, as far as our observations were restricted to the dispersed particles undergoing electrophoreses. The size distributions given in Figures 1 and 4 validate that our microscopy method for investigating extremely dilute suspensions is able to characterize the electric properties of submicron bubbles without forming micron aggregates during the LbL processes. As a consequence, comparison of the first and second charge reversals using microscopic electrophoresis experiments has revealed the peculiarity of SB systems more definitely than before.

Evaluating the number ratio of the added cations to the anions on the uncoated SB surfaces in the formation of the positively charged M-SBs, the ratio was enormously large and far beyond the stoichiometric condition of the unity ratio ($\alpha = 1$). In contrast, the double-layer deposition followed the conventional behavior in two respects. First, we obtained identical concentrations of $C_{cm}$ irrespective of the molecular weights for the same species of NaPSS. Second, the close monomer concentrations of the added polycations and polyanions implied that the charge reversal of the PE-SB particles occurred after the completion of the stoichiometric complexation of the cationic and anionic PEs, meaning that the conventional stoichiometry applies to the double-layer deposition.

These findings verified that the attractive electrostatic interactions played a secondary role in the formation of the first layer of polycations on the bare SBs. However, at the same time, our results demonstrated that one can manage to create encapsulated SBs without using additional materials or specific gases as long as a tremendous amount of excess polycations were added in the formation of the first layer. The results shown in Figure 3b further clarified that subsequent LbL coverage can be performed under normal conditions determined by the electrical stoichiometry. Coating SBs solely with polycations is a promising strategy to create smart agents of colloidal fine bubbles, because the exclusive use of polymers facilitates the attachment of targeting ligands in order not only to accomplish the site-specific delivery of drugs or genes, but also to provide the stimulus-responsive agents for ultrasound imaging [1–3].

**Acknowledgments**

We are grateful to Ryohei Yoshida for his experimental help.

**Supplementary Materials**

Movie S1 shows sequentially the electrophoretic migration of negatively charged bare SBs in parallel to the external electric field applied in the left direction on the movie, the reverse electrophoresis of positively charged M-SBs by adding 1.2 μM S-PLL at pH 7.1 under the same electric field, and the Brownian motion of M-SB particles, where both bare SBs and M-SBs are observed as white dots by dark-field microscopy.

**References**

1. Tsuge, H. *Micro-and Nanobubbles: Fundamentals and Applications*; CRC Press: Boca Raton, FL, USA, 2014.
2. Stride, E.; Edirisinghe, M. Novel microbubble preparation technologies. *Soft Matter* **2008**, *4*, 2350–2359.
3. Sirsi, S.R.; Borden, M.A. Microbubble compositions, properties and biomedical applications. *Bubble Sci. Eng. Technol.* **2009**, *1*, 3–17.
4. Borden, M. Nanostructural features on stable microbubbles. *Soft Matter* **2009**, *5*, 716–720.
5. Dressaire, E.; Bee, R.; Bell, D.C.; Lips, A.; Stone, H.A. Interfacial polygonal nanopatterning of stable microbubbles. *Science* **2008**, *320*, 1198–1201.
6. Lentacker, I.; De Smedt, S.C.; Sanders, N.N. Drug loaded microbubble design for ultrasound triggered delivery. *Soft Matter* **2009**, *5*, 2161–2170.
7. Duncan, P.B.; Needham, D. Test of the Epstein-Plesset model for gas microparticle dissolution in aqueous media: Effect of surface tension and gas undersaturation in solution. *Langmuir* **2004**, *20*, 2567–2578.
8. Takahashi, H.; Morita, A. A molecular dynamics study on inner pressure of microbubbles in liquid argon and water. *Chem. Phys. Lett.* **2013**, *573*, 35–40.
9. Shchukin, D.G.; Köhler, K.; Möhwald, H.; Sukhorukov, G.B. Gas-filled polyelectrolyte capsules. *Angew. Chem. Int. Ed.* **2005**, *44*, 3310–3314.
10. Winterhalter, M.; Sonnen, A.F.P. Stable air bubbles—catch them if you can! *Angew. Chem. Int. Ed.* **2006**, *45*, 2500–2502.
11. Lentacker, I.; De Geest, B.G.; Vandenbroucke, R.E.; Peeters, L.; Demeester, J.; De Smedt, S.C.; Sanders, N.N. Ultrasound-responsive polymer-coated microbubbles that bind and protect DNA. *Langmuir* **2006**, *22*, 7273–7278.
12. Borden, M.A.; Caskey, C.F.; Little, E.; Gillies, R.J.; Ferrara, K.W. DNA and polylysine adsorption and multilayer construction onto cationic lipid-coated microbubbles. *Langmuir* **2007**, *23*, 9401–9408.
13. Rossi, S.; Waton, G.; Krafft, M.P. Small phospholipid-coated gas bubbles can last longer than larger ones. *Chem. Phys. Chem.* **2007**, *9*, 1982–1985.
14. Lentacker, I.; De Smedt, S.C.; Demeester, J.; Van Marck, V.; Bracke, M.; Sanders, N.N. Lipoplex-loaded microbubbles for gene delivery: A Trojan horse controlled by ultrasound. *Adv. Funct. Mater.* **2008**, *17*, 1910–1916.
15. Park, J.I.; Tumarkin, E.; Kumacheva, E. Small, stable, and monodispersed bubbles encapsulated with biopolymers. *Macromol. Rapid Commun.* **2010**, *31*, 222–227.
16. Park, J.I.; Jagadeesan, D.; Williams, R.; Oakden, W.; Chung, S.; Stanisz, G.J.; Kumacheva, E. Microbubbles loaded with nanoparticles: A route to multiple imaging modalities. *ACS Nano* **2010**, *4*, 6579–6586.
17. Daiguji, H.; Matsuoka, E.; Muto, S. Fabrication of hollow poly-allylamine hydrochloride/poly-sodium styrene sulfonate microcapsules from microbubble templates. *Soft Matter* **2010**, *6*, 1892–1897.

18. Cornejo, J.J.M.; Matsuoka, E.; Daiguji, H. Size control of hollow poly-allylamine hydrochloride/poly-sodium styrene sulfonate microcapsules using the bubble template method. *Soft Matter* **2011**, *7*, 1897–1902.
19. Xu, Q.; Nakajima, M.; Liu, Z; Shiina, T. Biosurfactants for microbubble preparation and application. *Int. J. Mol. Sci*. **2011**, *12*, 462–475.
20. Nakatsuka, M.A.; Mattrey, R.F.; Esener, S.C.; Cha, J.N.; Goodwin, A.P. Aptamer-crosslinked microbubbles: Smart contrast agents for thrombin-activated ultrasound imaging. *Adv. Mater*. **2012**, *24*, 6010–6016.
21. Yap, R.K.; Whittaker, M.; Diao, M.; Stuetz, R.M.; Jefferson, B.; Bulmuş, V.; Peirson, W.L.; Nguyen, A.V.; Henderson, R.K. Hydrophobically-associating cationic polymers as micro-bubble surface modifiers in dissolved air flotation for cyanobacteria cell separation. *Water Res*. **2014**, *61*, 253–262.
22. Kovalenko, A.; Polavarapu, P.; Pourroy, G.; Waton, G.; Krafft, M.P. pH-controlled microbubble shell formation and stabilisation. *Langmuir* **2014**, *30*, 6339–6347.
23. Mahalingam, S.; Raimi-Abraham, B.T.; Craig, D.Q.; Edirisinghe, M. Formation of protein and protein–gold nanoparticle stabilized microbubbles by pressurized gyration. *Langmuir* **2014**, *31*,659–666.
24. Ohgaki, K.; Khanh, N.Q.; Joden, Y.; Tsuji, A.; Nakagawa, T. Physicochemical approach to nanobubble solutions. *Chem. Eng. Sci*. **2010**, *65*, 1296–1300.
25. Ushikubo, F.Y.; Furukawa, T.; Nakagawa, R.; Enari, M.; Makino, Y.; Kawagoe, Y.; Shiina, T.; Oshita, S. Evidence of the existence and the stability of nano-bubbles in water. *Colloids Surf. A Physicochem. Eng. Asp*. **2010**, *361*, 31–37.
26. Frusawa, H.; Inoue, M. When microbubble-polyelectrolyte complexes overcharge: A comparative study using electrophoresis. *Chem. Lett*. **2011**, *40*, 372–374.
27. Frusawa, H.; Yoshida, R. A non-stoichiometric universality in microbubble-polyelectrolyte complexation. *J. Phys. Soc. Jpn*. **2012**, *81*, SA008, doi:10.1143/JPSJS.81SA.SA008.
28. Liu, S.; Kawagoe, Y.; Makino, Y.; Oshita, S. Effects of nanobubbles on the physicochemical properties of water: The basis for peculiar properties of water containing nanobubbles. *Chem. Eng. Sci*. **2013**, *93*, 250–256.
29. Wu, C.; Nesset, K.; Masliyah, J.; Xu, Z. Generation and characterization of submicron size bubbles. *Adv. Colloid Interface Sci*. **2012**, *179*, 123–132.
30. Weijs, J.H.; Seddon, J.R.T.; Lohse, D. Diffusive shielding stabilizes bulk nanobubble clusters. *ChemPhysChem*. **2012**, *13*, 2197–2204.
31. Uchida, T.; Oshita, S.; Ohmori, M.; Tsuno, T.; Soejima, K.; Shinozaki, S.; Take, Y.; Mitsuda, K. Transmission electron microscopic observations of nanobubbles and their capture of impurities in wastewater. *Nanoscale Res. Lett*. **2011**, *6*, 1–9.
32. Agarwal, A.; Ng, W.J.; Liu, Y. Principle and applications of microbubble and nanobubble technology for water treatment. *Chemosphere* **2011**, *84*, 1175–1180.
33. Shen, Y.; Powell, R.L.; Longo, M.L. Interfacial and stability study of microbubbles coated with a monostearin/monopalmitin-rich food emulsifier and PEG40 stearate. *J. Colloid Interface Sci*. **2008**, *321*, 186–194.

34. Ikeura, H.; Hamasaki, S.; Tamaki, M. Effects of ozone microbubble treatment on removal of residual pesticides and quality of persimmon leaves. *Food Chem*. **2013**, *138*, 366–371.
35. Newman, C.M.H.; Bettinger, T. Gene therapy progress and prospects: ultrasound for gene transfer. *Gene Ther*. **2007**, *14*, 465–475.
36. Hernot, S.; Klibanov, A.L. Microbubbles in ultrasound-triggered drug and gene delivery. *Adv. Drug Deliv. Rev*. **2008**, *60*, 1153–1166.
37. Cavalieri, F.; Finelli, I.; Tortora, M.; Mozetic, P.; Chiessi, E.; Polizio, F.; Brismar, T.B.; Paradossi, G. Polymer microbubbles as diagnostic and therapeutic gas delivery device. *Chem. Mater*. **2008**, *20*, 3254–3258.
38. Tinkov, S.; Bekeredjian, R.; Winter, G.; Coester, C. Microbubbles as ultrasound triggered drug carriers. *J. Pharm. Sci*. **2009**, *98*, 1935–1961.
39. Yin, T.; Wang, P.; Li, J.; Zheng, R.; Zheng, B.; Cheng, D.; Li, R.; Lai, J.; Shuai, X. Ultrasound-sensitive siRNA-loaded nanobubbles formed by hetero-assembly of polymeric micelles and liposomes and their therapeutic effect in gliomas. *Biomaterials* **2013**, *34*, 4532–4543.
40. Yin, T.; Wang, P.; Zheng, R.; Zheng, B.; Cheng, D.; Zhang, X.; Shuai, X. Nanobubbles for enhanced ultrasound imaging of tumors. *Int. J. Nanomed*. **2012**, *7*, 895.
41. Suzuki, R.; Takizawa, T.; Negishi, Y.; Utoguchi, N.; Maruyama, K. Effective gene delivery with novel liposomal bubbles and ultrasonic destruction technology. *Int. J. Pharm*. **2008**, *354*, 49–55.
42. Brotchie, A.; Zhang, X.H. Response of interfacial nanobubbles to ultrasound irradiation. *Soft Matter* **2011**, *7*, 265–269.
43. Cavalli, R.; Bisazza, A.; Lembo, D. Micro-and nanobubbles: A versatile non-viral platform for gene delivery. *Int. J. Pharm*. **2013**, *456*, 437–445.
44. Kleimann, J.; Gehin-Delval, C.; Auweter, H.; Borkovec, M. Super-stoichiometric charge neutralization in particle-polyelectrolyte systems. *Langmuir* **2005**, *21*, 3688–3698.
45. Takahashi, M. ζ potential of microbubbles in aqueous solutions: Electrical properties of the gas-water interface. *J. Phys. Chem. B* **2005**, *109*, 21858–21864.
46. Oliveira, C.; Rubio, J. Zeta potential of single and polymer-coated microbubbles using an adapted microelectrophoresis technique. *Int. J. Miner. Process*. **2011**, *98*, 118–123.
47. Cohen, J.; Priel, Z.; Rabin, Y. Viscosity of dilute polyelectrolyte solutions. *J. Chem. Phys*. **1988**, *88*, 7111–7116.
48. Palberg, T.; Versmold, H. Electrophoretic-electroosmotic light scattering. *J. Phys. Chem*. **1989**, *93*, 5296–5301.
49. Ohshima, H.; Furusawa, K. *Electrical Phenomena at Interfaces: Fundamentals: Measurements, and Applications*; CRC Press: Boca Raton, FL, USA, 1998; Volume 76.